\documentclass[12pt]{article}
\begin{document}

\title{\bf \Large Effects of Force and Energy in an Evolving
Universe with a Gravitational Source}

\author{M. Sharif \thanks{e-mail: hasharif@yahoo.com}
\\ Department of Mathematics, University of the Punjab,\\ Quaid-e-Azam
Campus Lahore-54590, PAKISTAN.}

\date{}

\maketitle

\begin{abstract}
Two models are given by crossing the Friedmann metrics with
Schwarzschild and Kerr metrics. In these evolving universes with a
gravitational source, the force four-vector and the corresponding
potentials are evaluated.
\end{abstract}

\newpage


\section{Introduction}


The Friedmann models are without gravitational source due to their
homogeneity. There are models which combine the Friedmann universe
with a Schwarzschild metric [1], but only one metric acts at any
point in the spacetime. There being no model which could show both
the expansion of the universe and a gravitational attraction
together. Bokhari and Qadir [2] presented an alternative way of
constructing a toy model which gives an effect of a gravitational
source in an evolving universe. The spatial part of this metric is
the same as the Schwarzschild metric, but multiplied by a time
dependent scale factor. The time component is the usual
Schwarzschild time component

\begin{equation}
ds^2=e^{\nu(r)} dt^2-a^2(t)[e^{-\nu (r)}dr^2+r^2d\Omega^2],
\end{equation}
where $d\Omega^2=d\theta^2+\sin^2\theta d\phi$ is the solid angle.
This metric admits of a conformal time-like Killing vector. It
provides a gravitational source for the flat Friedmann model and
cannot be extended to the open and closed Friedmann universes. The
force was evaluated by using the conformal pseudo-Newtonian
$(c\psi N)$-formalism [2] which deals with conformally static
spacetimes. The time component of the force four-vector does not
appear there. Thus the metric and the $c\psi N$-formalism has
weakness of its own. In this paper we construct a new metric which
is applicable to all the three (flat, open, closed) Friedmann
models. Further we take another metric which is a cross between
the Kerr and the Friedmann metrics only for flat case. We evaluate
the force four-vector and energy imparted to a test particle for
these metrics so as to be able to analyse them in terms of forces
and energy. To this end we use the extended $(e\psi N)$-formalism
which deals with non-static spacetimes explicitly. We will not
discuss the $e\psi N$-formalism in any detail as that is discussed
separately [3,4].

The plan of the paper is as follows. In the next section we shall
briefly review the essential points of the $e\psi N$-formalism for
the purposes of application. In section three we determine the
forces and energy for the Friedmann-Schwarzschild crossed metric
and analyse them. In the next section we evaluate the $e\psi N$
force and potential for the Friedmann-Kerr crossed metric only for
the flat case. Finally in the fifth section the results are
summarised and discussed.

\section{The $e\psi N$-Formalism}

The basis of the formailsm is the observation that the tidal
force, which is operationally determinable, can be related to the
curvature tensor by

\begin{equation}
F_T^\mu=mR_{\nu\rho\pi}^\mu t^\nu l^\rho t^\pi,\quad (\mu,\nu,\rho,\pi=0,1,2,3),
\end{equation}
where $m$ is the mass of a test particle, $t^\mu=f(x)\delta_0^\mu,
\quad f(x)=(g_{00})^{-1/2}$ and $l^\mu$ is the separation vector.
$l^\mu$ can be determined by the requirement that the tidal force
have maximum magnitude in the direction of the separation vector.
Choosing a gauge in which $g_{0i}=0$ (similar to the synchronous
coordinate system [5,6]) in a coordinate basis. We further use
Riemann normal coordinates (RNCs) for the spatial direction, but
not for the temporal direction. The reason for this difference is
that both ends of the accelerometer are spatially free, i.e. both
move and do not stay attached to any spatial point. However, there
is a ``memory" of the initial time built into the accelerometer in
that the zero position is fixed then. Any change is registered
that way. Thus ``time" behaves very differently from ``space".

The $e\psi N$ force, $F_\mu$, satisfies the equation

\begin{equation}
F_T^{*\mu}=l^\nu F_{;\nu}^\mu ,
\end{equation}
where $F_T^{*\mu}$ is the extremal tidal force corresponding to
the maximum magnitude reading on the dial. Notice that
$F_T^{*0}=0$ does not imply that $F^0=0$. The requirement that
Eq.(3) be satisfied can be written as
\begin{equation}
l^i(F_{,i}^0+\Gamma_{ij}^0F^j)=0,
\end{equation}
\begin{equation}
l^j(F_{,j}^i+\Gamma_{0j}^iF^0)=F_T^{*i}.
\end{equation}

A simultaneous solution of the above equations can be found by taking the ansatz

\begin{equation}
F^0=m\left[(\ln A)_{,0}-\Gamma_{00}^0+\Gamma_{0j}^i\Gamma_{0i}^i/A\right]f^2,
\end{equation}
\begin{equation}
F^i=m\Gamma_{00}^i f^2,
\end{equation}
where $A=(\ln \sqrt{-g})_{,0},\quad g=det(g_{ij})$. These
equations can be written in terms of two quantities $U$ and $V$
given by

\begin{equation}
U=m\left[\ln (Af/B)-\int(g^{ij}_{,0} g_{ij,0}/4A)dt\right],
\end{equation}
\begin{equation}
V=-m\ln f,
\end{equation}
as
\begin{equation}
F_0=-U_{,0}, \quad F_i=-V_{,i}.
\end{equation}
It is to be noted that the momentum four-vector $p_\mu$ can be
written in terms of the integral of the force four-vector $F_\mu$.
Thus

\begin{equation}
p_{_\mu }=\int F_\mu dt.
\end{equation}
Notice that the zero component of the momentum four-vector
corresponds to the energy imparted to a test particle of mass $m$
while the spatial components give the momentum imparted to a test
particle.

\section{Friedmann-Schwarzschild Crossed Metric}

We define the metric by taking ``a cross" between the Friedmann
and Schwarzschild metrics by [4]

\begin{equation}
ds^2=e^{\nu(t,\chi)}dt^2-a^2(t)\left[e^{-\nu(t,\chi)}d\chi^2+\sigma^2(\chi)d\Omega^2\right],
\end{equation}
where $e^{\nu(t,\chi)}=[1-2M/a\sigma(\chi)], \chi$ is the
hyperspherical angle, $\sigma(\chi)$ is $\sinh\chi, \chi$ or
$\sin\chi$ according as the model is open $(k=-1),$ flat $(k=0)$
or closed $(k=1)$ and $a(t)$ is the corresponding scale parameter.
Even in the flat case, when $\sigma^2(\chi)=\chi^2$, Eq.(12) does
not reduce to Eq.(1) as the coefficient of $d\chi^2$ is time
dependent here. Since the physical distance is being re-scaled
this metric seems (relatively) more realistic than given by
Eq.(1). Ofcourse neither is a realistic cosmological model. Since
the conformal time-like Killing vector of the previous metric is
no longer available the $c\psi N$-formalism cannot now be applied.

The $e\psi N$ force, for the flat Friedmann model, is simply

\begin{equation}
\left.
\begin{array}{l}
F_0=\frac{3m\chi}{2(3\chi
a_0^{2/3}t^{2/3}-7Ma_0^{1/3})}[\frac{-4M^2a_0^{1/3}}{9\chi
t^{5/3}(a_0^{1/3}t^{2/3}-2M)}+\frac{2(a_0^{2/3}\chi-7Ma_0^{1/3}t^{-2/3})}{9\chi
t^{1/3}}\\
\qquad+\frac{4M(3a_0^{2/3}\chi-7Ma_0^{1/3}t^{-2/3})}{9\chi
(a_0^{1/3}t\chi-2Mt^{1/3})}],\\
F_1=-\frac{mM}{a_0^{1/3}t^{2/3}\chi^2(1-2M/a\chi)},\quad
F_2=F_3=0.
\end{array}
\right\}
\end{equation}

The time at which the repulsive force of the model inverts to an
attractive force can be obtained by making $F_0=0$. It will be

\begin{equation}
t_I=\frac{(M/2\chi)^{2/3}}{a_0^{1/2}}.
\end{equation}
We shall call this the ``inversion time".

For the open Friedmann model (for sufficiently small values of
$t$, for a given $\chi$), the $e\psi N$ force, is

\begin{equation}
\left.
\begin{array}{l}
F_0=-\frac{8m\sinh\chi\{(12t/a_0)^{-5/3}-\frac{1}{60}(a_0/12t)\}\{1-\frac{3}{20}
(12t/a_0)^{2/3}\}}{3a_0\sinh\chi\{\frac{1}{2}(12t/a_0)^{2/3}
+\frac{3}{40}(12t/a_0)^{4/3}\}-14M}[\frac{12M^2}{a_0\sinh\chi(a\sinh\chi-2M)}\\
\qquad\{1-\frac{3}{10}(12t/a_0)^{2/3}\}+\frac{7M}{a_0\sinh\chi}\{1+\frac{1}{30}
(12t/a_0)^{2/3}\}],\\
F_1=-\frac{4mM(a_0/12t)^{2/3}}{a_0\sinh^2\chi(1-2M/a\sinh\chi)}\{1-\frac{3}{20}
(12t/a_0)^{2/3}\},\quad
F_2=F_3=0.
\end{array}
\right\}
\end{equation}
The inversion time for the open Friedmann model will be

\begin{equation}
t_I=M/3\sinh\chi.
\end{equation}

For the closed Friedmann model of the universe (for sufficiently
small $t$, for the given $\chi$), this takes the form

\begin{equation}
\left.
\begin{array}{l}
F_0=-\frac{8m\sin\chi\{(12t/a_0)^{-5/3}+\frac{1}{60}(a_0/12t)\}\{1+\frac{3}{20}
(12t/a_0)^{2/3}\}}{3a_0\sin\chi\{\frac{1}{2}(12t/a_0)^{2/3}-\frac{3}{40}
(12t/a_0)^{4/3}\}-14M}[\frac{12M^2}{ a_0\sin\chi(a\sin\chi-2M)}\\
\qquad\{1+\frac{3}{10}(12t/a_0)^{2/3}\}+\frac{7M}{a_0\sin\chi}\{1-\frac{1}{30}
(12t/a_0)^{2/3}\}],\\
F_1=-\frac{4mM(a_0/12t)^{2/3}}{a_0\sin^2\chi(1-2M/a\sin\chi)}\{1+\frac{3}{20}
(12t/a_0)^{2/3}\},\quad F_2=F_3=0.
\end{array}
\right\}
\end{equation}
The inversion time for the closed model turns out to be

\begin{equation}
t_I=\frac{a_0}{2}\left[\pi/2+4M/3a_0\sin\chi-(8M/3a_0\sin\chi)^{1/2}-1\right].
\end{equation}

It is to be noted that the $e\psi N$ force, for the first order,
comes out to be equal for each of the Friedmann models of the
universe. The time component of the $e\psi N$ force, in each of
the Friedmann universe models, gives a measure of the change of
the gravitational potential energy of the test particle. The
spatial component represents a $\psi N$ force for the
Schwarzschild metric, modulo a local Lorentz factor for the flat
case. However, this component reduces for the early stages of the
open Friedmann model and increases for the early stages of the
closed Friedmann universe. Further, we see from Eqs.(15) and (17)
that the magnitude of the time component of the $e\psi N$ force
decreases for the early stages of the open Friedmann model while
it increases for the early stages of the closed Friedmann model.
The fact that we get the usual Newtonian force, as happens for the
Schwarzschild metric, shows that our metric does, infact, give the
effect of a gravitating particle of mass $m$. This is in agreement
with the already evaluated force for the Schwarzschild metric in
the $\psi N$ and the $e\psi N$-formalisms.

The $e\psi N$ potential for the flat Friedmann model will be
\begin{equation}
\left.
\begin{array}{l}
p_0=m\ln[\frac{3t(a_0^{1/3}t^{2/3}\chi-2M)^2}{2(a_0^{1/3}t^{2/3})^{12/7}
(3a_0^{1/3}t^{2/3}\chi-2M)^{9/7}}],\\
V=m\ln(1-2M/a_0^{1/3}t^{2/3}\chi)^{1/2}.
\end{array}
\right\}
\end{equation}
From here we note that $p_0$ tends to infinity as $t$ approaches
to zero. Infact, this gives the energy imparted to the test
particle in the Friedmann model for the flat case.

It is worth mentioning that ``in crossing" the two metrics the
potentials have merely been ``added". In principle each could have
been modified by the other as well.

For the open Friedmann universe, the $e\psi N$ potential is

\begin{equation}
\left.
\begin{array}{l}
p_0=m\ln[\frac{(a_0\sinh\chi)^{12/7}\{3a_0(\cosh\eta-1)\sinh\chi-4M\}^{9/7}\sinh\eta}
{a_0\{a_0(\cosh\eta-1)\sinh\chi-4M\}^2(\cosh\eta-1)^{2/7}}],\\
V=m\ln[\frac{a_0(\cosh\eta-1)\sinh\chi-4M}{a_0(\cosh\eta-1)\sinh\chi}]^{1/2}.
\end{array}
\right\}
\end{equation}

For the closed model of the Friedmann universe, the $e\psi N$
potential turns out to be

\begin{equation}
\left.
\begin{array}{l}
p_0=m\ln[\frac{(a_0\sin\chi)^{12/7}\{3a_0(1-\cos\eta)\sin\chi-4M\}^{9/7}\sin\eta}
{a_0\{a_0(1-\cos\eta)\sin\chi-4M\}^2(1-\cos\eta)^{2/7}}],\\
V=m\ln[\frac{a_0(1-\cos\eta)\sin\chi-4M}{a_0(1-\cos\eta)\sin\chi}]^{1/2}.
\end{array}
\right\}
\end{equation}

The quantity $p_0$ yields the energy of the test particle for this
crossed metric and the quantity $V$ gives the usual $\psi N$
potential for the Schwarzschild metric, modulo a local Lorentz
factor. It is worth noting that the time variaton and the usual
Newtonian gravitational potential are acting together in this
example. From here we see that the energy of the test particle
becomes infinite at time $t=0$ in each of the Friedmann models.

Notice that the force is repulsive at the early stages of the open
Friemann model of the universe. But after a particular time, the
attaractive force dominates the repulsive force. Thus there is an
attractive component of the cosmological force in the expanding
universe. This result coincides with the numerical results [7]
which also indicate the dominance of the attractive force.

\section{Friedmann-Kerr Crossed Metric}

We have obtained some fundamentally new insights by considering
the force four-vector for the Friedmann-Schwarzschild crossed
metric. Further insights can be expected from the force
four-vector by considering a more complicated metric than the
Friedmann-Schwarzschild crossed metric, namely a non-static
Kerr-like metric. This metric differs from the previous metric in
that the Friedmann-Schwarzschild crossed metric does not have any
conformal time-like Killing vector but it has a conformal
time-like Killing vector. This metric has been defined [9] by
multiplying the spatial part by a time factor. We shall call this
metric the Friedmann-Kerr crossed metric. This has the following
form

\begin{equation}
\left.
\begin{array}{l}
ds^2=(1-2Mr/R^2)dt^2-a^2(t)[(R^2/J)dr^2+R^2d\theta^2+(P\sin^2\theta/R^2)d\phi^2]\\
\qquad+\{2Mrba(t)\sin^2\theta/R^2\}dtd\phi,
\end{array}
\right\}
\end{equation}
where

\begin{equation}
R^2=r^2+b^2\cos^2\theta,\quad J=r^2-2Mr+b^2,\quad P=(r^2+b^2)^2-Jb^2\sin^2\theta
\end{equation}
and $b(=s/M)$ is the spin or angular momentum per unit mass of the
balck hole. The metric coefficients are given by

\begin{equation}
\left.
\begin{array}{l}
g_{00}=1-2Mr/R^2,\quad g_{11}=-a^2R^2/J,\quad g_{22}=-a^2R^2,\\
g_{33}=-a^2P\sin^2\theta/R^2,\quad
g_{03}=g_{30}=Mrab\sin^2\theta/R^2.
\end{array}
\right\}
\end{equation}

Under suitable coordinate transformations [6], the off-diagonal
elements vanish and we are left with the following metric
coefficients

\begin{equation}
\left.
\begin{array}{l}
g_{00}=g_{00}-g^{33}g_{03}^2=(1-2M/r)(1+4M^2b^2r^2\sin^2\theta/R^4J),\\
g_{11}=-a^2R^2/J,\quad g_{22}=-a^2R^2,\quad g_{33}=-a^2P\sin^2\theta/R^2.
\end{array}
\right\}
\end{equation}

The $e\psi N$ force, for the flat Friedmann model of the universe $(k=0)$ is

\begin{equation}
\left.
\begin{array}{l}
F_0=m/3t,\\
F_1=\frac{m}{JR^2(R^2-2Mr)(R^4J+4M^2b^2r^2\sin^2\theta)}[MR^4J^2(r^2-b^2\cos^2\theta)\\
\qquad+4M^2b^2\sin^2\theta\{R^2J(2r^3-3Mr^2+rb^2\cos^2\theta)\\
\qquad-r^2(R^2-2Mr)(3rJ+(r-M)R^2)\}],\\
F_2=\frac{mMrb^2\sin 2\theta}{R^2(R^2-2Mr)(R^4J+4M^2b^2r^2\sin^2\theta)}[R^4J-2Mr\{(R^2-2Mr)\\
\qquad(r^2+b^2+2b^2\sin^2\theta)-b^2\sin^2\theta\}],\\
F_3=0.
\end{array}
\right\}
\end{equation}

Notice that the zero component coincides with that of the zero
component of the flat Friedmann model [4]. It gives a rate of
change of energy which approaches infinity at the very early
stages of the Friedmann universe and goes to zero as $t$ tends to
infinity [4]. It is worth noting that the spatial components of
the $e\psi N$ force $F_1, F_2$ and $F_3$ are just the $\psi N$
force for the Kerr metric for a special choice of geodesics
[9,10], modulo a local Lorentz factor.

The $e\psi N$ ``potentials" for this spacetime turn out to be

\begin{equation}
\left.
\begin{array}{l}
U=-m\ln(t/T)-\frac{1}{2}m\ln[(R^2-2Mr)(R^4J+4M^2b^2r^2\sin^2\theta)/R^6J],\\
V=\frac{1}{2}m\ln[(R^2-2Mr)(R^4J+4M^2b^2r^2\sin^2\theta)/R^6J].
\end{array}
\right\}
\end{equation}
Here if we add both these potentials the resultant will be the
time component of the Friedmann metrics. Notice that the time
component of the $e\psi N$ potential comes out to be the time
component of the flat Friedmann model of the universe minus the
radial and polar coordinate dependent term. The additional term
occurs due to the $g_{00}$ of the Kerr metric. Thus the potentials
have merely been ``added" when we ``cross" the two metrics as in
the case of the previous metric.

The expressions for the $e\psi N$ force and the $e\psi N$
potential are comprehensible and help us in understanding the
energy of the test particle. We note that the spatial component of
the $e\psi N$ potential reduces to the usual $\psi N$ potential of
the Kerr metric for a special choice of geodesics [9,10], modulo a
local Lorentz factor. The sum of both these potentials gives the
potential energy imparted to the test particle. This energy goes
to infinity for $t=0$ as required for the Friedmann models of the
universe.

\section{Conclusion}

We have constructed a cross model which gives an effect of
gravitational source in all the Friedmann models of the universe.
We then applied the $e\psi N$-formalism to this crossed metric so
as to obtain the physical effects of forces. The spatial component
of the force four-vector and the scalar quantity V give the
Newtonian gravitational force and potential respectively for each
of the Friedmann universes. This shows that the effect of
gravitational source in an evolving universe. We have seen that
the inversion time is different for each of the Friedmann models.
Further we attempted to use the Kerr metric instead of
Schwarzschild metric only for the flat case. In this case the time
component just gives the energy imparted to a test particle for
the flat Friedmann universe. We get physically acceptable effects
in terms of forces and energy in each of the Friedmann models. The
problem of constructing a new metric (i.e. cross between Kerr and
Friedmann models) for each of the Friedmann metrics and then
applying the $e\psi N$-formalism to it remains open.

\newpage

\begin{description}
\item {\bf Acknowledgments}
\end{description}

The author would like to thank Prof. Chul H. Lee for his
hospitality at the Department of Physics and Korea Scientific and
Engineering Foundation (KOSEF) for postdoc fellowship at Hanyang
University Seoul, KOREA.

\vspace{2cm}

{\bf \large References}

\begin{description}

\item{[1]} Lindquist R.W. and Wheeler J.A.: Rev. Mod. Phys. {\bf
29}(1957)432;\\ Einstein A. and Strauss E.: Rev. Mod. Phys. {\bf
17}(1945)120; {\bf 18}(1946)148.

\item{[2]} Bokhari A.H.: Ph.D. Thesis (Quaid-i-Azam University, 1985);\\
Bokhari A.H. and Qadir A.: {\it Proc. 4th Marcel Grossmann Meeting
on General Relativity}, ed. R. Ruffini (Elsevier Science
Publishers, 1986) 1635.

\item{[3]} Qadir, A. and Sharif, M.: {\it Proc. 4th Regional Conference on
Mathematical Physics}, eds. F. Ardalan, H. Arfaei and S. Rouhani
(Sharif University of Tech. Press, 1990);\quad Nuovo Cimento B{\bf
107}(1992)1071.

\item{[4]} Sharif M.: Ph.D. Thesis (Quaid-i-Azam University, 1991).

\item{[5]} Misner, C.W., Thorne, K.S. and Wheeler, J.A.: {\it Gravitation}
(W.H. Freeman San Francisco, 1973).

\item{[6]} Landau, L.D. and Lifschitz, E.M.: {\it The Classical Theory of
Fields} (Pergamon Press, 1975).

\item{[7]} Qadir, Asghar and Siddiqui, A.W.: work in progress.

\item{[8]} Bokhari, A.H.: Nuovo Cimento B{\bf 103}(1989)617.

\item{[9]} Qadir, A. and Quamar, J.: {\it Proc. 4th Marcel Grossmann Meeting
on General Relativity}, ed. Hu Ning (Science Press and North
Holland Co., 1983)189;\\ Quamar, J.: Ph.D. Thesis (Quaid-i-Azam
University, 1984).

\item{[10]} Qadir, A. and Quamar, J.: Europhys. Lett. {\bf 4}(1986)423.

\end{description}

\end{document}